\documentclass[preprint,12pt]{elsarticle}

\usepackage{amssymb}
\usepackage{amsthm}
\usepackage{amsmath}
\usepackage{mathabx}
\usepackage{xcolor}
\usepackage{nicefrac}
\usepackage{hyperref}
\usepackage{multirow}
\usepackage{makecell}
\usepackage{units}

\newcounter{bla}

\def\k{{\bf k}}
\def\r{{\bf r}}
\def\L{{\bf L}}
\def\volr{{\Omega_{\mathrm{cell}}}}

\journal{Computer Physics Communications}

\begin{document}

\begin{frontmatter}

\title{DensityTool: A post-processing tool for space- and spin-resolved density of states from VASP}

\author[a]{Lucas Lodeiro\corref{author}}
\author[b]{Tom\'{a}\v{s} Rauch\corref{author}}
\cortext[author] {Corresponding author.\\\textit{E-mail address:} lucas.lodeiro@ug.uchile.cl , tomas.rauch@uni-jena.de}
\address[a]{Departamento de Qu\'imica, Facultad de Ciencias, Universidad de Chile, Las Palmeras 3425, \~Nu\~noa 7800003, Santiago, Chile}
\address[b]{Institut f\"{u}r Festk\"{o}rpertheorie und -Optik, FSU Jena, Max-Wien-Platz 1, 07743 Jena, Germany, Germany}

\begin{abstract}
The knowledge of the local electronic structure of heterogeneous solid materials is crucial for understanding their electronic, magnetic, transport, optical, and other properties. VASP, one of the mostly used packages for density-functional calculations, provides local electronic structure either by projecting the electronic wave functions on atomic spheres, or as a band-decomposed partial charge density. Here, we present a simple tool which takes the partial charge density and the energy eigenvalues calculated by VASP as input and constructs local charge and spin densities. The new data provides a much better spatial resolution than the projection on the atomic spheres. It can be visualized directly in the real space e.g. with Vesta, or averaged along planes spanned by two of the lattice vectors of the periodic unit cell. The plane-averaged local (spin) density of states can be easily plotted e.g. as color-coded data using almost any plotting program. DensityTool can be applied to manipulate, visualize, and understand the local electronic structure of any system calculated with VASP. We expect it to be useful especially for researchers concerned with inhomogeneous systems, such as interfaces, defects, surfaces, adsorbed molecules, or hybrid inorganic-organic composites.

\end{abstract}


\end{frontmatter}

{\bf PROGRAM SUMMARY}

\begin{small}
\noindent
{\em Program Title:} DensityTool \\
{\em CPC Library link to program files:} (to be added by Technical Editor) \\
{\em Developer's repository link:} \url{https://github.com/llodeiro/DensityTool}\\
{\em Code Ocean capsule:} (to be added by Technical Editor) \\
{\em Licensing provisions:} MIT \\
{\em Programming language:} FORTRAN \\
{\em Supplementary material:} A complementary User Manual and examples can be found in the developer's repository link. \\
{\em Nature of problem(approx. 50-250 words):} 
Total and local density of states (LDOS) are widely-used quantities for understanding the electronic structure of condensed matter. In VASP, one of the mostly used packages for numerically solving the Kohn-Sham equations of density-functional theory, one does not directly obtain the LDOS. Usually, one works with partial density of states (PDOS), where the wave function is projected onto atomic orbitals. Alternatively, it is possible to calculate the band- and wavevector-dependent charge densities. Together with the band structure, LDOS can be obtained using post-processing tools.\\

{\em Solution method(approx. 50-250 words):} 
DensityTool combines the band structure with band- and wavevector-dependent charge densities calculated by VASP to construct LDOS. In addition, spin-resolved LDOS (LSDOS) can be calculated for magnetic systems. Thus obtained LDOS and LSDOS can be further processed with DensityTool. In particular, it can be averaged along chosen spatial directions, which is convenient for example to demonstrate the spatially-resolved electronic structure of inhomogeneous systems.\\
DensityTool can handle all three types of approaches implemented in VASP (non-magnetic, collinear magnetic, non-collinear magnetic including spin-orbit coupling), as well as all types of self-consistent band calculations as DFT (LDA, GGA, mGGA and hybrid functionals) and HF.\\
For convenience, the output of DensityTool follows the format of VASP (the CHGCAR file), which can be easily used in other programs for visualization. We provide a sample python script to plot averaged L(S)DOS from the calculated data. \\

{\em Additional comments including restrictions and unusual features (approx. 50-250 words):}    \\
   \\
\end{small}

\section{Introduction}
Density-functional theory (DFT)~\cite{Hohenberg1964,Kohn1965} is an extremely successful approach to calculating ground-state electronic properties of condensed matter in the one-particle approximation~\cite{https://doi.org/10.1002/andp.19273892002}. In this work we will refer to applications of DFT to periodic systems, which allows to use the implications of the Bloch theorem~\cite{Bloch29}. The systems are described in terms of periodic unit cells, which may be also of artificial nature. This allows to describe with this approach also non-periodic materials like finite slabs with surfaces, 2D materials, or molecules.

For a given periodic system, the Kohn-Sham~\cite{Hohenberg1964,Kohn1965} equations are solved with the wave functions $\varphi_{n,\k}(\r)$ and their eigenvalues $\epsilon_{n,\k}$ as a solution. The square of the occupied wave functions yields the partial charge density $P_{n,\k}(\r) = \left| \varphi_{n,\k}(\r) \right|^2$ and the eigenvalues are interpreted as the one-particle energy levels.

Using the energy levels, it is common to define the density of states (DOS) as
\begin{equation}
    D(E) = \frac{N_e}{(2\pi)^3} \sum_n \int_{\mathrm{BZ}} \delta(E-\epsilon_{n,\k})\ d^{3}k
\end{equation}
with $N_e$ the band occupancy. The DOS is an important property which describes the complete electronic structure of the entire material. This is very convenient in case of periodic solids with rather smaller unit cells, where the spatial variation of DOS is not important.

On the other hand, when one is studying the electronic structure of inhomogeneous systems like e.g. interfaces, defects, surfaces, adsorbed molecules, or hybrid inorganic-organic composites, access to spatially resolved DOS is more convenient. Such a quantity is the local DOS (LDOS) which combines the energy levels with the partial charge density:
\begin{equation}
    \label{eq:LDOS}
    L(E,\r) = \frac{N_e}{(2\pi)^3} \sum_n \int_{\mathrm{BZ}} \delta(E-\epsilon_{n,\k})P_{n,\k}(\r)\ d^{3}k.
\end{equation}
Alternatively, LDOS can be understood as energy-resolved partial charge density, starting with the definition of the total charge density
\begin{equation}
\rho(\r)=\frac{N_e\volr}{(2\pi)^3}\sum_n \int_{\mathrm{BZ}}f_{n,\k}P_{n,\k}(\r)\ d^{3}k
\end{equation}
with the unit cell volume $\volr$ and the Fermi-Dirac distribution function $f_{n,\k}$.

In magnetic systems the spin becomes an additional quantum number when it is conserved (e.g., for collinear magnetism and systems without spin-orbit coupling). We thus have the spin-resolved quantities
\begin{equation}
P^{\updownarrows}_{n,\k}(\r)=\left|\varphi^{\updownarrows}_{n,\k}(\r)\right|^2,
\end{equation}
\begin{equation}
\rho^{\updownarrows}(\r)=\frac{\volr}{(2\pi)^3}\sum_n \int_{\mathrm{BZ}}f^{\updownarrows}_{n,\k}P^{\updownarrows}_{n,\k}(\r)\ d^{3}k,
\end{equation}
\begin{equation}
    D^{\updownarrows}(E) = \frac{1}{(2\pi)^3} \sum_n \int_{\mathrm{BZ}} \delta(E-\epsilon^{\updownarrows}_{n,\k})\ d^{3}k,
\end{equation}
and
\begin{equation}
    \label{eq:LDOS_updown}
    L^{\updownarrows}(E,\r) = \frac{1}{(2\pi)^3} \sum_n \int_{\mathrm{BZ}} \delta(E-\epsilon^{\updownarrows}_{n,\k})P^{\updownarrows}_{n,\k}(\r)\ d^{3}k.
\end{equation}
Using these, the total quantities can be recovered as the sum of the spin-resolved parts, e.g., the total charge density becomes
\begin{equation}
\rho (\r) = \rho^{\uparrow} (\r) + \rho^{\downarrow} (\r).
\end{equation}

In addition, information about the spin-part of the magnetization can be deduced from the spin density
\begin{equation}
s (\r) = \rho^{\uparrow} (\r) - \rho^{\downarrow} (\r)
\end{equation}
and its spatially-resolved generalization, the local spin DOS (LSDOS)
\begin{equation}
    \label{eq:LSDOS}
    S(E,\r) = \frac{1}{(2\pi)^3} \sum_n \int_{\mathrm{BZ}}\left[ \delta(E-\epsilon^{\uparrow}_{n,\k})P^{\uparrow}_{n,\k}(\r) - \delta(E-\epsilon^{\downarrow}_{n,\k})P^{\downarrow}_{n,\k}(\r)\right]d^{3}k.
\end{equation}
LSDOS can be used to study the energy- and position-dependent contributions to the spin magnetization. For example, such an approach can be applied to identify the origin of local magnetization at surfaces~\cite{GRADMANN1991481,SurfMag} or defects~\cite{6566103,PhysRevLett.106.087205}.

In the following we present a tool which takes $P_{n,\k}$ and $\epsilon_{n,\k}$ calculated with the VASP~\cite{vasp3,Kresse1996,Kresse1999} package as input and calculates $L(E,\r)$ and $S(E,\r)$. Furthermore, the functionality of the tool allows to calculate $L$ and $S$ averaged along some direction for a better interpretation of the data.

\section{Methodology}
\label{sec:theory}
After successfully finishing a self-consistent calculation with the VASP package, the eigenstates $\epsilon_{n,\k}$ are listed in the EIGENVAL file. Setting further \texttt{LPARD=T, LSEPB=T, LSEPK=T} in the input file (INCAR), the partial charge densities $P_{n,\k}$ are calculated by VASP from the preconverged wave functions and stored in the PARCHG.nnnn.kkkk files. These files are the basic input of DensityTool.

In VASP, the structure of the calculated system is defined by a periodic unit cell spanned by the three lattice vectors $\L_i$, $i=1,2,3$. The volume of the unit cell $\Omega_{\mathrm{cell}} = L_1 L_2 L_3 |\hat{L}_3 \cdot (\hat{L}_1 \times \hat{L}_2)|$ with $L_i$ and $\hat{L}_i$ the magnitude and the unit vector of the lattice vectors $\L_i$. In the numerical calculation, the partial charge density (along with other quantities) is sampled on a discrete mesh in the real space with $N_i$ mesh points along each of the three lattice vectors and the eigenvalues (band structure) are sampled on another discrete mesh in the Brillouin zone in the reciprocal space. The spacing of both meshes can be set in VASP input files. Therefore, the L(S)DOS is also expressed on the real-space discrete mesh. To evaluate the Brillouin-zone integrals above, their discretized version
\begin{equation}
    \label{eq:BZ_approx}
    \frac{1}{(2\pi)^3} \int_{\mathrm{BZ}} \ldots\ d^{3}k
    \qquad \longrightarrow \qquad{} \frac{1}{\Omega_{\mathrm{cell}}}\sum_{\k\ \in\ \mathrm{BZ}}\ldots W_{\k}
\end{equation}
must be employed. $W_\k$ are weights of the $\k$-points and their values depend on the discretization of the reciprocal unit cell in the original VASP calculation.

With DensityTool we are aiming at the construction of the L(S)DOS and its visualization. Therefore, it is useful to express the $E$-dependence of $L$ and $S$ on a regular mesh, as it is already the case for the spatial coordinates. For this purpose, we approximate $\delta(E-\epsilon_{n,\k})$ in all equations by
\begin{equation}
    \label{eq:delta_approx}   
    \delta(E-\epsilon_{n,\k}) \longrightarrow g_{n,\k}=\frac{1}{\sigma \sqrt{2\pi}}e^{-\frac{(E-\epsilon_{n,\k})^2}{2\sigma^2}}
\end{equation}
with a proper smearing parameter $\sigma$. By this approximation $L(E,\r)$ and $S(E,\r)$ becomes continuous in $E$ and in particular it can be evaluated on an equidistant set of $E$ values. We further introduce a threshold parameter $\eta$ and redefine Eq.~\eqref{eq:delta_approx} as
\begin{equation}
\label{eq:gauss_fin}
 \tilde{g}_{n,\k}(E,\sigma) = \left\{\begin{array}{cl}
        g_{n,\k}(E,\sigma) & \text{for } g_{n,\k}(E,\sigma) > \frac{\eta}{\sigma\sqrt{2\pi}}\\
        0 & \text{for } g_{n,\k}(E,\sigma) \leq \frac{\eta}{\sigma\sqrt{2\pi}}.\\
        \end{array}\right.
\end{equation}
Thus, for a given energy $E$, only those $P_{n,\k}$ are considered, for which $g_{n,\k}(E,\sigma) > \frac{\eta}{\sigma\sqrt{2\pi}}$. This procedure ensures a substantial speed up of the summation over all $n$ and $\k$. As an optimal compromise we propose to set $\eta = 0.001$, for which $99.98\%$ of the distribution area of $g_{n,\k}$ is recovered.

Finally, DensityTool can also calculate the L(S)DOS averaged on a plane spanned by two of the lattice vectors for each of the discrete mesh points along the third lattice vector and for each $E$. This way three-dimensional data is obtained which can be visualized e.g. as a scatter plot with color coded $z$-coordinate of the plot. For example, we choose to average the data on the plane spanned by $\L_1$ and $\L_2$ with area $A_{12} = L_1 L_2 |\hat{L}_1 \times \hat{L}_2|$. We can thus define the plane-averaged LDOS as
\begin{equation}
    \label{eq:ldos_avg}
    \bar{L}(E,r_3) = \frac{1}{A_{12}}\iint L(E,\r) |\hat{L}_1\times \hat{L}_2| dr_1dr_2.
\end{equation}
For our numerical evaluation it is more convenient to rewrite Eq.~\eqref{eq:ldos_avg} as
\begin{equation}
    \label{eq:LDOS_avg_fin}
    \bar{L}(E,r_3) = \frac{N_e}{(2\pi)^3} \sum_n \int_{\mathrm{BZ}} \delta(E-\epsilon_{n,\k})\bar{P}_{n,\k}(\r)\ d^{3}k,
\end{equation}
where $\bar{P}_{n,\k}(r_3)$ is the partial charge average (PCA) over the area $A_{12}$,
\begin{equation}
\label{eq:pca}
\bar{P}_{n,\k}(r_3)= \frac{1}{A_{12}}\iint P_{n,\k}(\r) |\hat{L}_1\times \hat{L}_2| dr_1dr_2 = \frac{1}{L_1L_2}\iint P_{n,\k}(\r)dr_1dr_2.
\end{equation}
Note that the real-space integrals must be evaluated on the discrete mesh and thus
\begin{equation}
    \frac{1}{A_{12}} \iint \ldots |\hat{L}_1\times \hat{L}_2| dr_1dr_2\qquad\longrightarrow\qquad{} \frac{1}{N_1N_2}\sum_{r_1,r_2}\ldots.
\end{equation}

Using Eq.~\eqref{eq:pca} plane-averaged LSDOS or total spin or charge density can be calculated in analogy to the plane-averaged LDOS.

In this section we demonstrated how plane-averaged and total L(S)DOS is calculated from the output of VASP with DensityTool. A final subtlety remains related to magnetic systems with conserved (collinear) spin (\texttt{ISPIN=2} in VASP). In this case, the primary output of VASP is $P_{n,\k}(\r)$ and $s_{n,\k}(\r)=P^{\uparrow}_{n,\k}(\r) - P^{\downarrow}_{n,\k}(\r)$. The spin-resolved partial charge density $P^{\updownarrows}_{n,\k}(\r)$ can be obtained from this data as $P^{\updownarrows}_{n,\k}(\r) = \frac{1}{2}\left( P_{n,\k}(\r) \pm s_{n,\k}(\r) \right)$. In addition, we provide an alternative approach, which is computationally faster, but only applicable to cases with small spin splitting of the energy levels. In this approximation we use Eq.~\eqref{eq:LDOS} to calculate LDOS, but we have to approximate $\epsilon_{n,\k} \approx \frac{1}{2}(\epsilon^{\uparrow}_{n,\k}+\epsilon^{\downarrow}_{n,\k})$. LSDOS can be approximately calculated using the same approach, but by substituting $P_{n,\k}(\r)$ for $s_{n,\k}(\r)$ in Eq.~\eqref{eq:LDOS}.

\section{Software structure and workflow}
DensityTool is written in FORTRAN. The code is provided at \url{https://github.com/llodeiro/DensityTool}, including installation and execution instructions. It is a post-processor to VASP and it takes its output written to the EIGENVAL, CHGCAR, and PARCHG files as input. In addition, input parameters are necessary, which control the exact routines that should be used and the numerics. Most convenient way to do this is to execute DensityTool by letting it read the input parameters from a text file (see DENSITYTOOL.IN in our examples). Please find the list of the available input parameters in the User Manual.

Depending on the choice of routines, different tasks can be performed by DensityTool. For plane-averaged L(S)DOS, the PCA or partial spin average (PSA) must be calculated first with Eq.~\eqref{eq:pca}. For each state characterized by $(n,\k)$, the plane-averaged quantity is written to an individual file. This enables to conduct further calculations with much smaller files than the original ones written by VASP. The main tasks are then to calculate the plane-averaged L(S)DOS from the previously calculated PCA or PSA via Eq.~\eqref{eq:LDOS_avg_fin}. In case of magnetic materials, it is possible to choose between the exact treatment of the spin-split energy levels and the approximation described in Sec.~\ref{sec:theory}. Alternatively, it is also possible to calculate the full L(S)DOS without performing the average (Eqs.~\eqref{eq:LDOS} and \eqref{eq:LSDOS}). Again, in case of magnetic materials the user can choose to apply the approximation to the eigen-energies. Finally, the plane-averaged total charge or spin density can be calculated directly from the CHGCAR file written by VASP.

We summarize the available routines in Tab.~\ref{tab:routines} and provide their detailed description in the User Manual.
\begin{table}[ht]
    \centering
    \begin{tabular}{c|cc|cc}
            & \multicolumn{2}{c|}{\makecell{nonmagnetic or magnetic \\ with approximated energies}} & \multicolumn{2}{c}{\makecell{magnetic with \\exact energies}} \\
            \hline 
            & partial charge & partial spin & partial charge & partial spin \\
          \hline
        partial average & PCA & PSA & \multicolumn{2}{c}{PARCHGSPIN}\\
            \hline 
        \makecell{plane-averaged \\  local density} & LDOS & LSDOS & LDOSMAG & LSDOSMAG \\
            \hline         
        \makecell{full local \\ density} & LDOSFULL & LSDOSFULL & LDOSFULLMAG & LSDOSFULLMAG \\
            \hline         
        total density & \multicolumn{2}{c|}{CHGCARAVG} & \multicolumn{2}{c}{CHGCARSPIN}\\
                    \hline 
    \end{tabular}
    \caption{Routines available in DensityTool. The two columns refer to the use of the approximation to the spin-split energy levels for magnetic materials with conserved spin.}
    \label{tab:routines}
\end{table}

\section{Results}
We demonstrate the capabilities of DensityTool on the example of concrete inhomogeneous systems. This type of systems can be seen as a prototypical field where the tool can be successfully applied. We provide additional simple examples at the code repository \url{https://github.com/llodeiro/DensityTool}. To keep the amount of data low for these simple examples, the VASP calculations were not converged with respect to the number of $\k$-points in these examples, making the results unphysical and only suitable for testing. Therefore, we do not show these results here.
Note that there are also published results obtained with DensityTool in its development version~\cite{Rauch2021,interface}.

In particular, we present our results calculated with DensityTool for two inhomogeneous systems: a composite inorganic-organic system, and a perovskite slab model.

\subsection{Composite inorganic-organic system:}
\begin{figure}[ht!]
  \centering
  \includegraphics[width = 0.99\columnwidth]{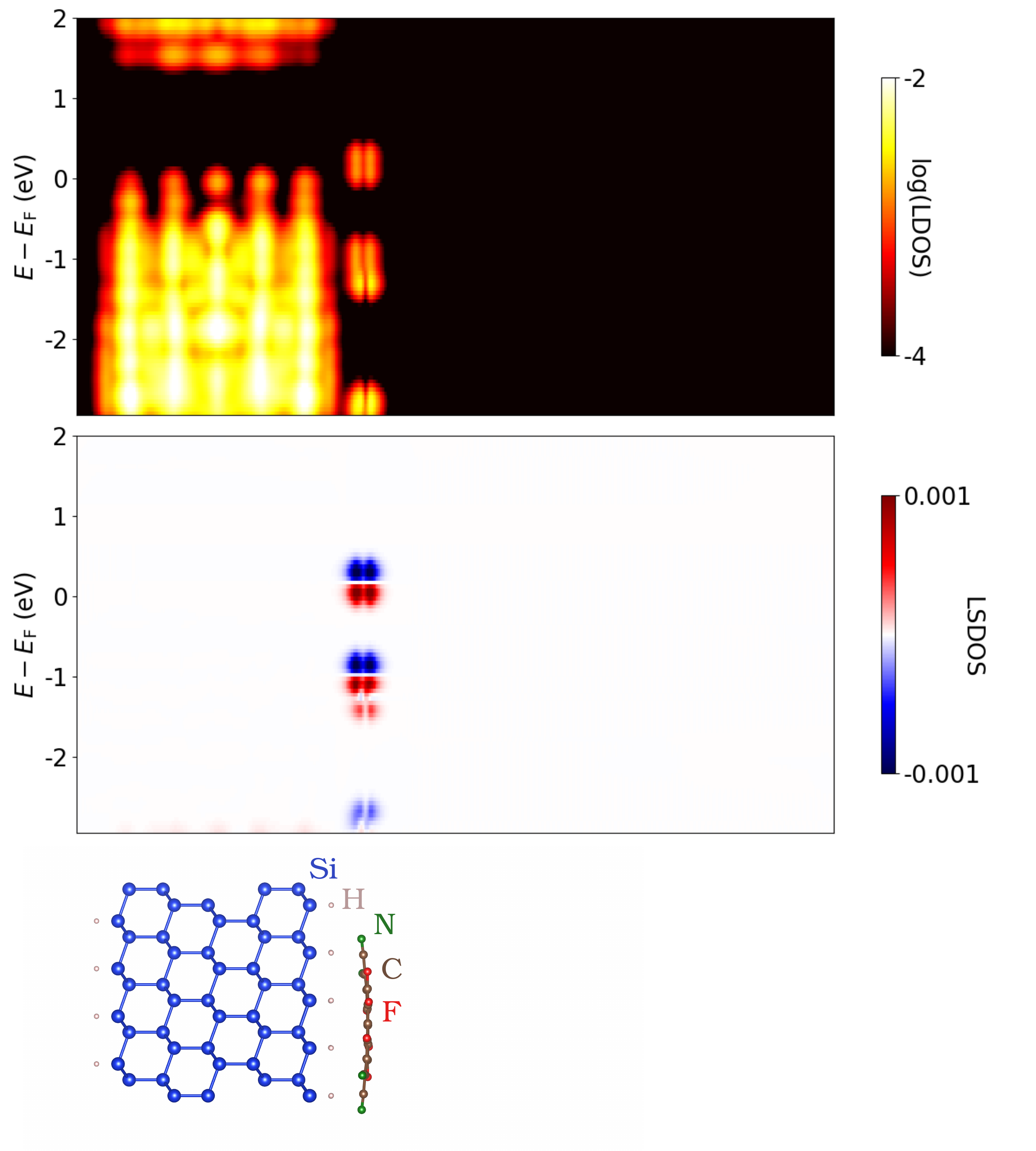}
  \caption{LDOS (top), LSDOS (center), and the atomic structure (bottom) of the hydrogenized Si(111) slab with the F6-TCNNQ molecule adsorbed to its surface. Some of the data shown here was published previously in Ref.~\cite{Rauch2021}. The figures were created with Matplotlib v2.2.2, VESTA v3.4.6~\cite{vesta}, and Inkscape 0.92.}
  \label{fig:ldos_Si_F6}
\end{figure}

The composite inorganic-organic system of interest is a hydrogenized Si(111) slab with the F6-TCNNQ molecule adsorbed to one of the surfaces. The atomic structure of the calculated system is shown in the bottom panel of Fig.~\ref{fig:ldos_Si_F6}. Find the details of the VASP electronic structure calculations in Ref.~\cite{Rauch2021} and more information about the system in Ref.~\cite{doi:10.1002/aelm.201800891}.
Using DensityTool, we calculated the LDOS and LSDOS of the system. We chose the exact treatment of the spin-resolved energies. Thus, starting from the partial charge density written by VASP, it was averaged in the plane parallel to the Si surface for both spin orientations separately with the PARCHGSPIN routine. Subsequently, the plane-averaged LDOS and LSDOS were calculated with the LDOSMAG and LSDOSMAG routines, respectively. We present the results in the top (LDOS) and central (LSDOS) panel of Fig.~\ref{fig:ldos_Si_F6}. It can be clearly seen that the plane-averaged LDOS has particularly large contributions in the regions of the Si atomic planes both below the Fermi energy $E_F$ (occupied states) and above the insulating band gap (unoccupied states). The location of the molecular orbitals of the F6-TCNNQ molecule both in real space as well as on the energy axis can be made visible with this method. In addition, the plane-averaged LSDOS gives insight into the magnetic properties of the system. While there is no magnetization visible for the Si states, the orbitals of the attached molecule appear to be spin-polarized. Note that this is in agreement with Fig.~4c of Ref.~\cite{doi:10.1002/aelm.201800891} and Fig.~4 of Ref.~\citep{Rauch2021}.
\begin{figure}[ht!]
  \centering
  \includegraphics[trim=900 350 900 300, clip, width = 0.49\columnwidth]{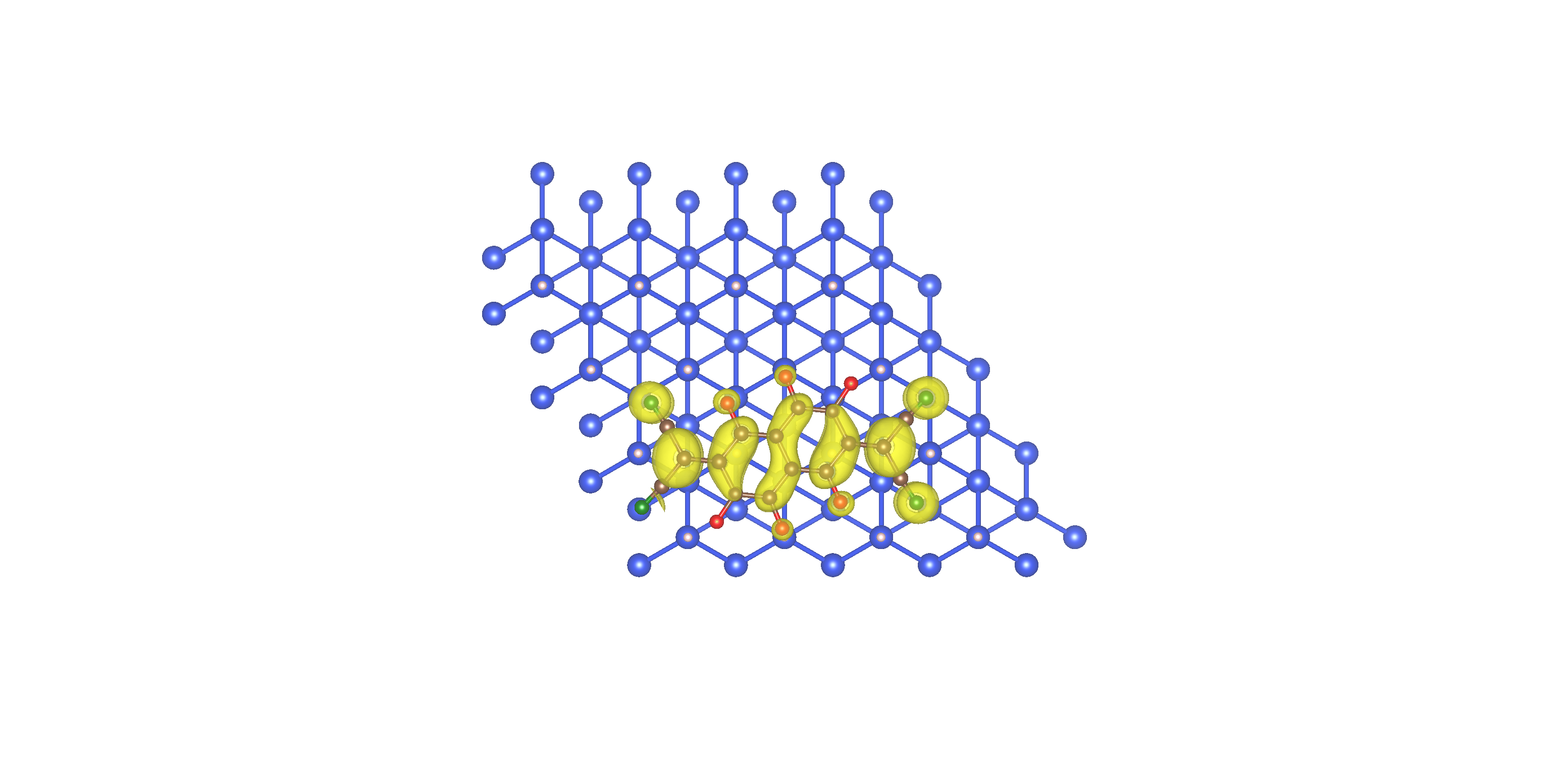}
  \caption{LSDOS (top view) of the hydrogenized Si(111) slab with the F6-TCNNQ molecule adsorbed to the surface as a constart-value contour at $E=E_F+$\unit[0.05]{eV}. The data (yellow) is shown as a contour plot. The color code of the atoms is the same as in the bottom panel of Fig.~\ref{fig:ldos_Si_F6}. The figure was created with VESTA 3.4.6~\cite{vesta}.}
  \label{fig:lsdos}
\end{figure}

Complementary to the plane-averaged data, we also calculated the full LSDOS using the LSDOSFULLMAG routine of DensityTool. We then set the energy to $E=E_F+$\unit[0.05]\ {eV} and visualized the LSDOS as a contour surface of constant spin polarization shown in Fig.~\ref{fig:lsdos}. In agreement with the central panel of Fig.~\ref{fig:ldos_Si_F6}, the LSDOS is significantly positive for the chosen energy only in the region around the molecule.

\subsection{Perovskite slab model:}
\begin{figure}[ht!]
  \centering
  \includegraphics[width = 0.99\columnwidth]{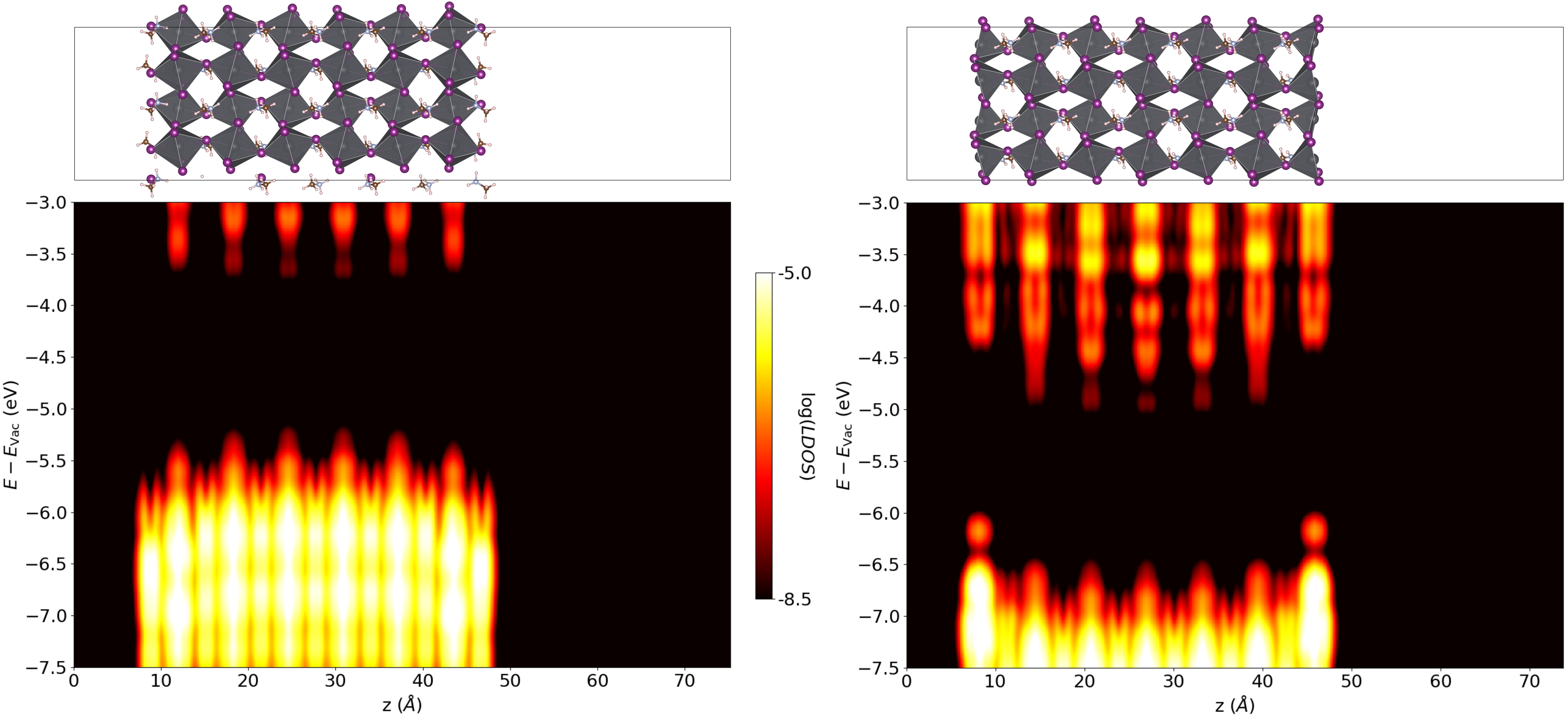}
  \caption{LDOS of MAI- (left) and PbI$_2$(\textit{unpolarized})-terminated (right) surface slabs. The atomic structure of each slab is added on top of each LDOS figure as a visual guide. The figures were created with Matplotlib v2.2.5, VESTA v3.4.3~\cite{vesta} and Inkscape 0.92.3.}
  \label{fig:LDOS_MAPI}
\end{figure}

In this example we studied a slab model of the CH$_3$NH$_3$PbI$_3$ perovskite, also known as MAPI. Particularly, we used ($2\times 2)\times6$ slabs of (001) MAI- and PbI$_2$(\textit{unpolarized})-terminated surfaces of MAPI, see Ref.~\cite{slabs} for slab nomenclature. Electronic structure was calculated with the HSE06 hybrid functional~\cite{hse,hse06} and spin-orbit coupling was included. For more calculation details see the supporting information in Ref. \cite{slabs}. The atomic structure for both slabs is shown on top of Fig.~\ref{fig:LDOS_MAPI}.

Using DensityTool, we calculated the LDOS of both systems in the non-collinear setting. For this case, the partial charge density written by VASP is the starting point. Only its average in the plane parallel to the slab surface is needed and it is calculated with the PCA routine. Subsequently, the plane-averaged LDOS is calculated around the band gap with the LDOS routine. We present the results in the left and right panel of Fig.~\ref{fig:LDOS_MAPI} for MAI- and PbI$_2$-terminated slabs, respectively.
From both plane-averaged LDOS profiles, it is evident that the electronic structure is highly correlated with the structure of alternated MAI and PbI$_2$ layers of this material. Particularly, the inner slab (bulk-like) states near the band gap (valence state maximum and conduction states) are composed mainly of PbI$_2$ layers. The PbI$_2$-terminated slab exhibits surface states located at the surface PbI$_2$ layers, which decrease the band gap of the entire system. This does not affect the inner slab (bulk-like) band gap, which is almost the same as the inner band gap of MAI-terminated slab. Finally, PbI$_2$-terminated slab states are remarkably deeper in energy \mbox{($\sim$1.3 eV)}, which means the magnitude of the ionization potential of this exposed surface is larger than for the MAI-terminated one. These conclusions are consistent with Ref.~\cite{slabs}.

\section{Conclusion}
With DensityTool we provide a practical post-processor to VASP, which allows for the manipulation and visualization of the partial charge density in the form of LDOS or LSDOS, which can be also averaged in a plane spanned by two of the three lattice vectors of the periodic unit cell. This approach is complementary and in many cases superior to the commonly used approach of projecting the DOS on the atomic spheres. Using the partial charge density and DensityTool, a better spatial resolution can be achieved and also vacuum regions can be studied. The L(S)DOS in real space can be also visualized as a constant-value contour for chosen energy values.

DensityTool takes directly the output of VASP as input. The plane-averaged L(S)DOS is written to a set of files (one for each energy value) and it can be plotted with any common plotting program. Full L(S)DOS is written in the VASP format (e.g. the one of CHGCAR) for compatibility. The data can thus be easily visualized using e.g. VESTA~\cite{vesta}.

We believe that DensityTool will be utilized to improve the understanding of the local electronic structure of inhomogeneous systems, such as interfaces, defects, surfaces, adsorbed molecules, or hybrid inorganic-organic composites.

\section*{Declaration of competing interest}
The authors declare that they have no known competing financial interests or personal relationships that could have appeared to influence the work reported in this paper.

\section*{Acknowledgment}

We thank Prof. Silvana Botti, Prof. Eduardo Men\'{e}ndez-Proupin, and Prof. Francisco Mu\~noz for support and fruitful discussions. T. Rauch acknowledges funding from the Volkswagen Stiftung (Momentum) through the project ``dandelion''. 



\bibliographystyle{elsarticle-num}
\bibliography{main.bbl}

\end{document}